\documentclass[11pt]{article}
\usepackage{moriond,epsfig}

\usepackage{amsmath}
\usepackage{amsfonts}
\usepackage{amssymb}

\def\be{\begin{equation}}
\def\ee{\end{equation}}
\def\bea{\begin{eqnarray}}
\def\eea{\end{eqnarray}}

\begin{document}
\vspace*{4cm}
\title{STRONG ELECTROWEAK SYMMETRY BREAKING}

\author{ANNA KAMI\'{N}SKA}

\address{Institute of Theoretical Physics, Faculty of Physics, \\
University of Warsaw, Ho{\.z}a 69, 00-681, Warsaw, Poland}

\maketitle\abstracts{
While the LHC takes on the challenge of experimentally exploring the electroweak symmetry breaking sector, it is not only interesting but also crucial to explore alternatives to the Standard Model scenario with an elementary scalar Higgs boson. The idea of electroweak symmetry breaking by some new strong dynamics is discussed. A simple, general and self-consistent low energy effective description of Higgsless models is introduced. This effective theory is studied from the point of view of prolonging perturbative unitarity of $WW$ scattering by spin-1 resonances originating from the strongly interacting sector. The LHC phenomenology and the discovery potential for these spin-1 resonances is also discussed. The role of spin-1 resonances is then considered on the grounds of composite Higgs models. A general prescription for the explicit inclusion of such resonances in the effective Lagrangian description of these models is presented.}

\section{Introduction}

The only missing and so far undiscovered element of the Standard Model is the mechanism of spontaneous $SU(2)_{L}\times U_{Y}$ symmetry breaking, known also as the Higgs mechanism. Even though we do not know what is the new sector which realizes electroweak symmetry breaking, we can learn something about it by pointing out minimal ingredients which it should provide
\begin{enumerate}
\item three Goldstone bosons to become longitudinal degrees of freedom of $W$ and $Z$; such Goldstone bosons should originate from spontaneous breaking of some global symmetry and should have appropriate quantum numbers; the Brout-Englert-Higgs mechanism is then used to trade these unphysical massless bosons for longitudinal polarizations of gauge bosons, which become massive,
\item most probably - the custodial $SU(2)_{C}$ symmetry which would protect the value of the $\rho$ parameter,
\item new degrees of freedom prolonging perturbative unitarity of $WW$ scattering amplitudes; without any new degrees of freedom contributing to scattering amplitudes of longitudinally polarized $W$ and $Z$ bosons, perturbative unitarity condition for these amplitudes becomes violated at the energy scale of $\sim 1.5\ TeV$.
\end{enumerate} 
The minimal sector based on scalar fields which provides these ingredients is the SM Higgs sector. Here we will explore a class of non-standard electroweak symmetry breaking models, where the symmetry becomes broken by some new strong dynamics. The main motivation for this idea is that symmetry breaking by strong interactions is observed in Nature (for example in QCD), while no elementary scalar boson has been observed so far. If the electroweak symmetry is broken by strong dynamics, a totally new strongly interacting sector is needed. What is however crucially important from the point of view of experimentally exploring the new strongly interacting sector, is whether we can provide an effective perturbative description of its effects at low energies. As one might expect, the key to solving this problem can be found in our experience with QCD. Chiral perturbation theory offers a systematic effective framework for investigating strong interaction processes at low energies. The effective Lagrangian is based on derivative expansion and describes in a perturbative regime interactions of composite states connected with some symmetry breaking pattern by treating them as fundamental degrees of freedom. In fact the same global symmetry breaking pattern $SU\left( 2\right) _{L} \times SU\left( 2\right) _{R}\ \rightarrow\ SU\left( 2\right) _{V}$ which generates QCD pions as pseudo-Goldstone bosons could produce in our new strongly interacting sector exactly the desired amount of Goldstone bosons required for $W^{\pm}_{L}$ and $Z_{L}$ (which would have a composite nature). With the global $SU(2)_{V}$ left unbroken this general effective description of the new strongly interacting sector provides already two out of three minimal ingredients outlined in this section. At last we can expect that the scattering amplitudes of longitudinally polarized $W$ and $Z$ bosons at energies $m_{W}^{2}\ll s\ll \Lambda_{strong}^{2}$ will be unitarized by effects of the new strongly interacting sector which can be described effectively in terms of weakly interacting resonances.

Again from the point of view of exploring the strongly interacting electroweak symmetry breaking sector experimentally, the most important are the lightest, lowest-lying resonances which give a dominating contribution to scattering amplitudes of longitudinal $W$ and $Z$ bosons. There are two most 'natural' candidates for the lowest-lying resonance
\begin{enumerate}
\item spin-1 resonances; they are motivated by QCD, where the scattering amplitudes of pions are saturated by the $\rho$ meson triplet (the scalar $\sigma$ resonance plays a negligible role). Vector meson dominance in $W_{L}W_{L}$ scattering amplitudes is predicted by a large group of specific models incorporating strong electroweak symmetry breaking. These models are described often as Higgsless models. Two large and most known families of Higgsless models are related to technicolor and deconstruction of extra dimensions. Though Higgsless models are often challenged by electroweak precision data, they provide many interesting scenarios of strong electroweak symmetry breaking.
\item composite scalar; in order to make the composite Higgs of the new strongly interacting sector naturally lighter than the scale of strong dynamics, it is often introduced as a pseudo-Goldstone boson of an enlarged global symmetry group. Composite Higgs models are more naturally in consistence with electroweak precision data than Higgsless models and they do not suffer under the hierarchy problem of the Standard Model.
\end{enumerate}
It is useful to note that whatever the nature of the lowest-lying resonance really is, this resonance does not fully unitarize the $W_{L}W_{L}$ scattering amplitudes. Even if the lightest physical degree of freedom originating from the strong sector is the composite scalar, one would expect to observe spin-1 resonances at some higher energies. Hence the existence of spin-1 resonances seems to be one of the most distinctive features of strong electroweak symmetry breaking models from the point of view of their low-energy phenomenology.

\section{Higgsless models}

\begin{figure}[h!]
\begin{minipage}[b]{0.47\linewidth}
\centering
\includegraphics[width=8 cm]{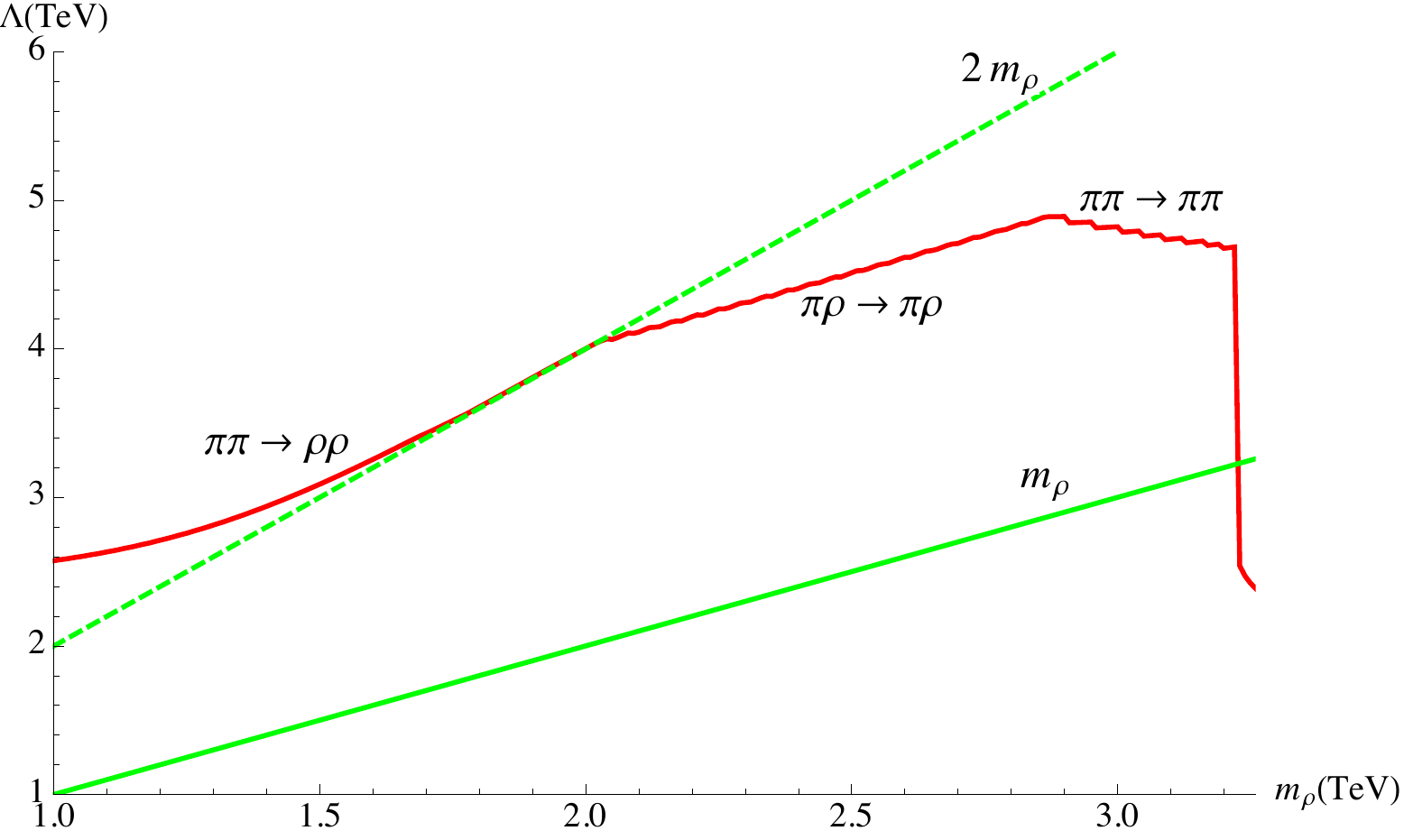}
\caption{Maximal possible cutoff $\Lambda$ as a function of $m_{\rho}$ resulting form the sum of all considered perturbative unitarity conditions.}
\label{SUMdocP}
\end{minipage}
\hspace{0.5cm}
\begin{minipage}[b]{0.47\linewidth}
\centering
\includegraphics[width=7 cm]{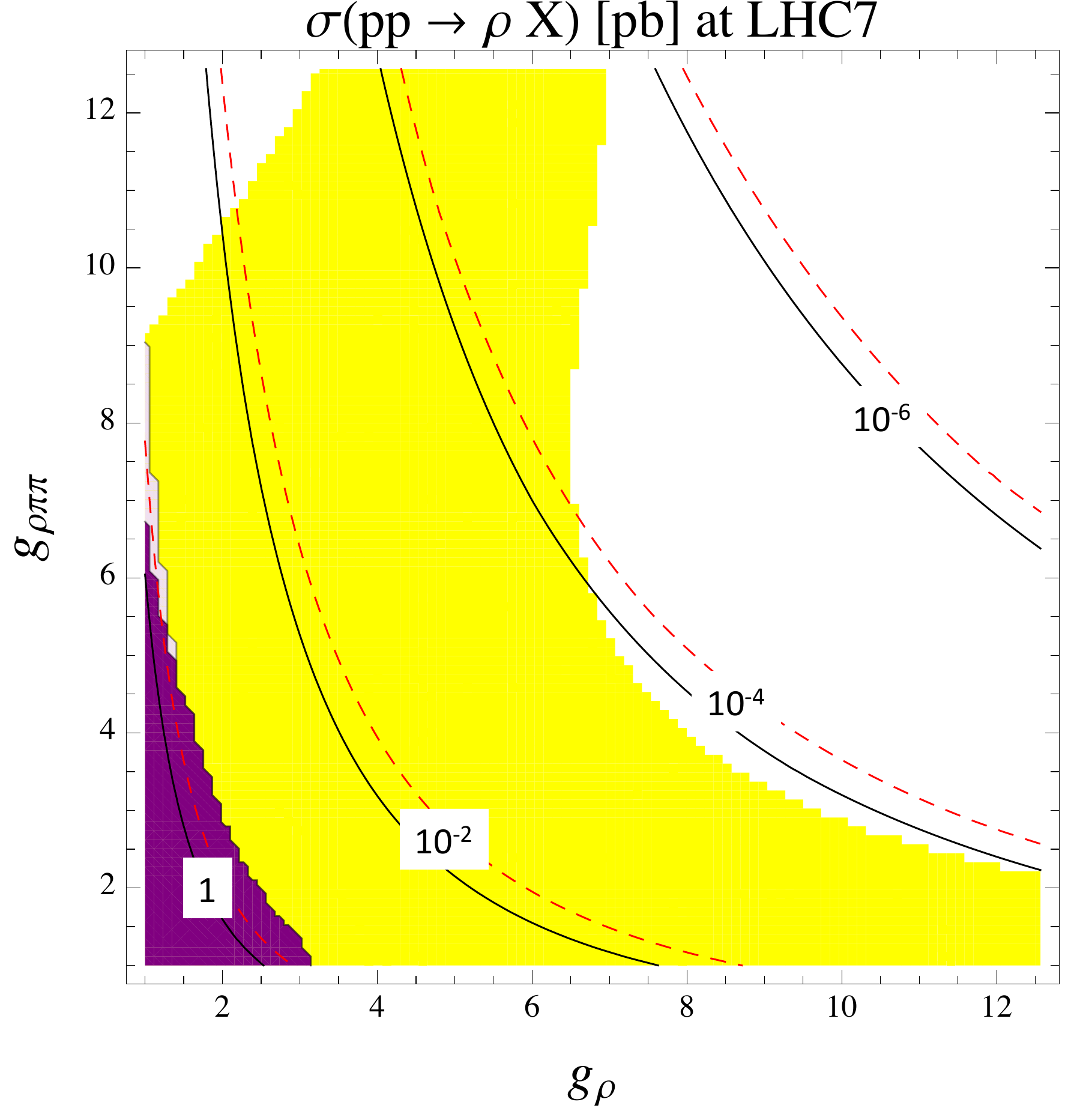}
\caption{Contours of the inclusive $\rho^{0},\ \rho^{\pm}$ (normal and dashed lines) production cross section at the LHC with $\sqrt{s}=7\ TeV$}
\label{newlimits1}
\end{minipage}
\end{figure}

In this section \cite{Falkowski:2011ua} we provide a simple, general and self-consistent effective low energy framework to study Higgsless models, which provide a perfect laboratory for studying the physics of spin-1 resonances. In the effective description we explicitly introduce the lowest-lying spin-1 resonances into the effective Lagrangian, which is built under the following minimal assumptions
\begin{itemize}
\item vector meson dominance, which means that all the couplings of the effective Lagrangian will be saturated by effects of the lightest spin-1 resonances
\item approximate global $SU(2)_{C}$.
\end{itemize}
We use the 'hidden gauge' formalism \cite{Bando:1987br}, which has proven to describe the properties of $\rho$ mesons in QCD very well. The entire construction is based on the
\begin{equation}
\mathcal{G}=SU\left( 2\right) _{L} \times SU\left( 2\right) _{R}\ \rightarrow\ \mathcal{H}=SU\left( 2\right) _{h} 
\end{equation}
symmetry breaking pattern which produces the minimal required number of Goldstone bosons and leaves a global $SU(2)$ custodial symmetry unbroken. The idea of the 'hidden gauge' formalism is that we can factorize the matrix $U$ known from chiral perturbation theory in terms of two special unitary matrices $\xi_{L}$ and $\xi_{R}$ as follows
\begin{equation}
U=\xi_{L}U_{0}\xi_{R}^{\dag},\ \ U_{0}=\left\langle U\right\rangle  =1_{2\times 2},\ \ U \rightarrow g_{L}Ug_{R}^{\dag},\ \ g_{L,\; R}\in SU(2)_{L,\; R}
\end{equation}
where $U_{0}$ has been introduced as the $SU(2)_{L}\times SU(2)_{R}\rightarrow SU(2)_{h}$ breaking VEV. The new $\xi_{L,\; R}$ fields transform as 
\begin{equation}
\xi_{L,\; R} \rightarrow g_{L,\; R}\xi_{L,\; R}h^{\dag},\ \ \ \ \ \ \ h\in SU(2)_{h}
\end{equation}
under the global symmetry. Spin-1 resonances are introduced these fields into the effective Lagrangian by gauging $SU(2)_{h}$ and identifying $\rho_{\mu}$ as gauge bosons of the 'hidden' local symmetry. Couplings of the SM fields to the spin-1 resonances are introduced by covariant derivatives of the $\xi_{L,\; R}$ fields
\begin{eqnarray}
D_\mu \xi_L &=& \partial_\mu \xi_L - i  \frac{g}{2}  W_\mu^a \sigma^a  \xi_L  + i \frac{g_\rho  }{ 2} \xi_L \rho_\mu^a \sigma^a   \nonumber \\
D_\mu \xi_R &=& \partial_\mu \xi_R - i  \frac{g' }{ 2}  B_\mu  \sigma^3 \xi_R  + i  \frac{g_\rho }{ 2}  \xi_R \rho_\mu^a \sigma^a .
\end{eqnarray}
where $g_{\rho}$ is the 'hidden gauge' coupling. Building blocks of the effective Lagrangian are objects with definite parity, transforming in the adjoint representation of $SU(2)_{h}$
\begin{equation}
V_\mu^{\pm} =  \xi_L^\dagger  D_\mu \xi_L \pm \xi_R^\dagger D_\mu \xi_R .
\end{equation}
Then the most general parity preserving Lagrangian built out of $\xi_{L,\; R}$ at the leading order in derivative expansion takes the form
\begin{equation}
\label{e.LeadingLagrangianPC}
 \mathcal{L}^{\left( 2\right) }= - \frac{v^2 }{ 4}  \mathrm T \mathrm r \left\lbrace    \alpha V_\mu^+ V_\mu^+  +  V_\mu^- V_\mu^-   \right\rbrace .
\end{equation}
The Lagrangian has two free parameters $g_{\rho}$ and $\alpha$. Classical technicolor copying the structures of QCD corresponds to $\alpha\approx 2$ while the three-site model of dimensional deconstruction corresponds to $\alpha=1$. A careful derivation of physical degrees of freedom of the effective Lagrangian \ref{e.LeadingLagrangianPC} shows that, assuming the hierarchy $g,\; g'\ll g_{\rho}$, the original SM $W^{\pm}$ fields correspond in the first approximation to the lightest mass eigenstates of the effective Lagrangian, while the 'hidden gauge' $\rho^{\pm}$ field corresponds at the leading order in $g/g_{\rho}$ expansion to the heavy resonance with mass $m_{\rho}^2\approx \alpha g_\rho^2 v^2$. Also the original $\pi^{\pm}$ and $G^{\pm}$ Goldstone fields correspond in the first approximation to Goldstone eigenstates which are defined as fields which become traded for the longitudinal degrees of freedom of massive gauge boson eigenstates. The coupling of two pions (related to longitudinal $W$ and $Z$ bosons) to the $\rho$ resonance, which is relevant for deriving $WW$ scattering amplitudes using the Goldstone equivalence theorem, is given by $g_{\rho\pi\pi} = \frac{\alpha}{2}g_{\rho}$. Knowing the couplings of pions with spin-1 resonances we analyze perturbative unitarity in elastic $\pi\pi$ scattering, inelastic $\pi\pi$ scattering (including the $\pi\pi\rightarrow\rho\rho$ channel) and $\pi\rho$ scattering. The maximal possible cutoff resulting form the sum of all considered perturbative unitarity conditions as a function of $m_{\rho}$ is shown in Fig. \ref{SUMdocP}. One can see that the unitarity constraint from the $\pi\pi\rightarrow\rho\rho$ channel dominates for low values of $m_{\rho} \sim 1 − 2\; TeV$, placing the cutoff almost immediately above $2m_{\rho}$. For intermediate values of $m_{\rho} \sim 2.5\; TeV$ the most stringent constraint comes from the $\pi\rho\rightarrow\pi\rho$ channel, bringing the cutoff below $2m_{\rho}$. For large values of $m_{\rho} \sim 3\; TeV$ the $\pi\rho\rightarrow\pi\rho$ channel constraint becomes weaker while the elastic $\pi\pi\rightarrow\pi\pi$ channel determines the cutoff (which is below $2m_{\rho}$, so the $\pi\pi\rightarrow\rho\rho$ channel does not constrain it due to kinematical reasons). At $m_{\rho} \sim 3.2\; TeV$ the maximal cutoff drops rapidly because of the perturbativity constraint that we impose on resonance couplings. We can conclude that if a spin-1 resonance plays a dominant role in unitarizing $W_{L}W_{L}$ scattering amplitudes, it should be observed at energies below (or maximally of the order of) $3\; TeV$. We can also see that, counter-intuitively, a heavy $\rho$ meson of mass $\sim \left( 2.5-3\right) \; TeV$ is more efficient in prolonging perturbative unitarity than a light resonance of mass $\sim 2\; TeV$.

In order to say something about the LHC phenomenology we need to define the couplings of spin-1 resonances with matter fields. In the following we assume that the SM quarks and leptons are fundamental - they couple to the heavy resonances only via mixing of the latter with the SM gauge bosons. This mixing arises due to non-diagonal entries in the gauge boson mass matrix implied by the effective Lagrangian. At the leading order in $1/g_\rho$ the mass eigenstates are reached by the rotation of the SM gauge bosons
\begin{eqnarray}
W_\mu^\pm &\rightarrow &  W_\mu^\pm   -\frac {g}{2 g_\rho}   \rho_\mu^\pm , \nonumber \\ 
Z_\mu &\rightarrow & Z_\mu - \frac{ g^2 - g'{}^2}{2 g_\rho \sqrt{g^2+  g'{}^2} }\rho_\mu^0 ,  \nonumber \\
A_\mu &\rightarrow &  A_\mu - \frac{e}{2 g_\rho}  \rho_\mu^0 , 
\end{eqnarray}
and the corresponding rotation  of $\rho$. Under such assumptions the main discovery channel at the Tevatron and LHC is the search for resonant production of $W^+ W^-$ and $W^\pm Z$ pairs
\begin{equation}
\Gamma(\rho^0 \rightarrow  W^+ W^-)  \approx   \Gamma(\rho^\pm \rightarrow  Z W^\pm) \approx 
 \frac{m_\rho g_{\rho \pi \pi}^2}{48 \pi}  =  {m_\rho^5 }{192 \pi g_\rho^2 v^4}  \, . 
\end{equation}

\begin{figure}[h!]
\begin{minipage}[b]{0.47\linewidth}
\centering
\includegraphics[width=80 mm]{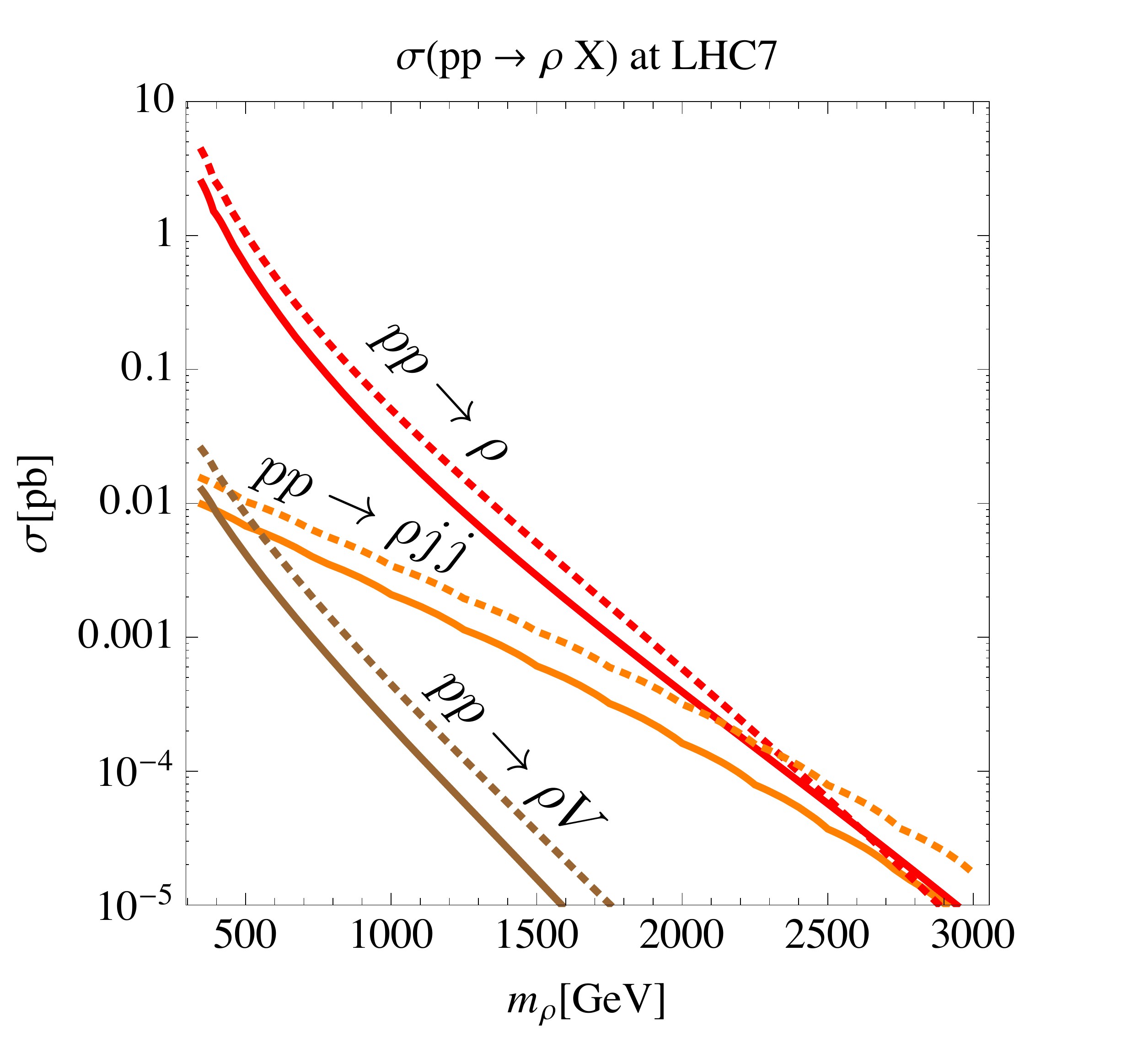}
\caption{Cross section for the production of a single neutral (solid) and charged (dashed) resonance at the LHC  with $\sqrt{s} = 7$ TeV. We set $g_\rho = 4$.}
\label{xsections_lhc7}
\end{minipage}
\hspace{0.5cm}
\begin{minipage}[b]{0.47\linewidth}
\centering
\includegraphics[width=80 mm]{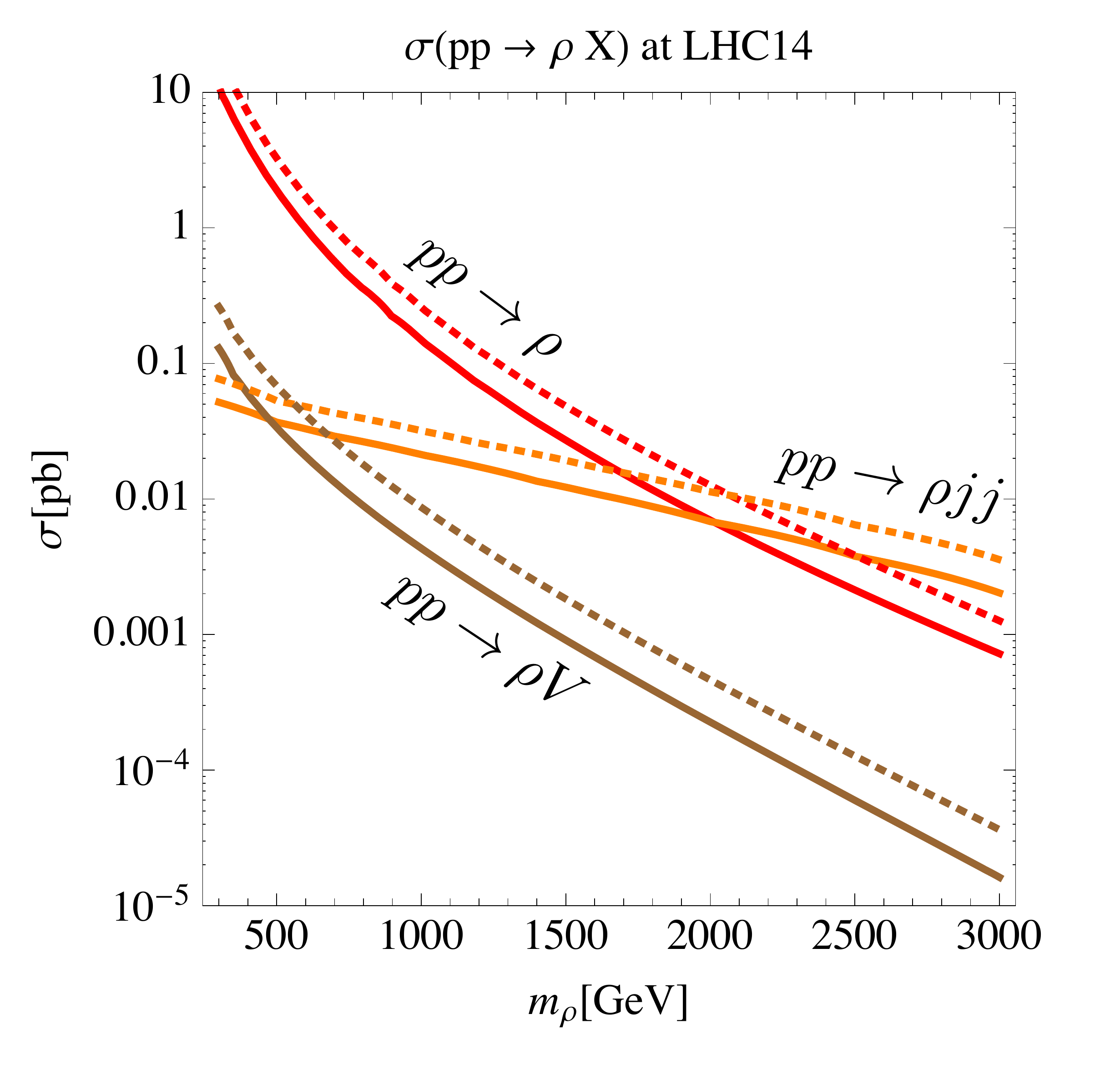}
\caption{Cross section for the production of a single neutral (solid) and charged (dashed) resonance at the LHC with $\sqrt{s} = 14$ TeV. We set $g_\rho = 4$.}
\label{xsections_lhc14}
\end{minipage}
\end{figure}

At the LHC the resonances are produced mainly via the following processes:
\begin{itemize}
\item Drell-Yan ($q \bar q \to \rho$)
\item Vector boson fusion (VBF) ($VV \to \rho$)
\item  $\rho-$strahlung ($V \to \rho V$).
\end{itemize} 
The production cross section depends on $m_\rho$ via the parton distribution functions. Furthermore, since the coupling of the resonances to the SM is suppressed by $1/g_\rho$ for a fixed $m_\rho$ the cross section for all the above processes scale as $1/g_{\rho}^2$. In Fig. \ref{xsections_lhc7} and Fig. \ref{xsections_lhc14} we plotted the cross sections for the 3 channels mentioned above at the LHC with $\sqrt{s} = 7$ TeV and $\sqrt{s} = 14$ TeV. The Drell-Yan process dominates in most of the parameter space. The VBF is suppressed by the 3-body final state phase space, however it becomes important for very heavy resonances, $m_\rho \gtrsim 2$ TeV. This is due to the fact that this process, unlike the two others, can be initiated by a quark-quark collision, and the quark-quark luminosity at the LHC decreases less rapidly than the quark-antiquark one. In Fig. \ref{newlimits1} we show the contours of the inclusive $\rho$ production cross section at the LHC with $\sqrt{s}=7\ TeV$ in the parameter space of $g_{\rho}$ and $g_{\rho\pi\pi}$. The cross sections were computed at tree level using the MSTW 2008 PDFs \cite{Martin:2009iq}. The yellow shape in the background of the plot shows the region of the parameter space allowed by perturbative unitarity. The purple shape shows the region of parameter space already excluded by the CMS search for WZ resonant production \cite{CMS}. One can deduce from this plot that if the resonances are heavy ($m_{\rho} \gtrsim 2\ TeV$) and strongly coupled they might escape any direct detection at the LHC.

\section{Composite Higgs models}

Let us now say a few words about composite Higgs models, where the composite scalar is naturally lighter than other resonances because it emerges as a pseudo Nambu-Goldstone (pNG) boson of an enlarged global symmetry of the strong dynamics. The existence of a light scalar state similar to the SM Higgs seems to be the easiest and most natural solution consistent with electroweak precision data. The minimal composite Higgs model \cite{Agashe:2004rs} in which only one physical composite scalar remains in the spectrum is given by the following global symmetry breaking pattern
\begin{equation}
\mathcal{G}\; =\; SO(5)\ \ \rightarrow\ \ \mathcal{H}\; =\; SO(4) .
\end{equation}
As the composite Higgs does not fully unitarize the $WW$ scattering amplitudes, the existence of other resonances with different spins is expected. The effects of these resonances might provide the only experimentally accessible way of testing the nature of the Higgs if its interactions are similar to the ones predicted by the Standard Model. In order to study the effects of these heavier resonances it is best to include them explicitly in the effective Lagrangian, be done for example by generalizing the 'hidden gauge' formalism to non-standard global symmetry breaking patterns \cite{Piai:2004yb}.

\section{Conclusions}

We have studied the physics of spin-1 resonances connected with strong EW symmetry breaking in a simple, general effective framework built using tools known from QCD (CHPT, 'hidden gauge'). Considering perturbative unitarity in this simple, general setup with spin-1 resonances constrains the allowed resonance mass and its couplings. A crucial role is played by inelastic scattering effects - as it turns out, a heavy $\rho$ meson (2.5-3) TeV is more efficient in prolonging perturbative unitarity than a light resonance ($\sim$2 TeV). If the resonances are heavy ($m_{\rho}\geq 2TeV$) and strongly coupled they might escape any direct detection at the LHC. It is interesting to use such effective frameworks to study strong electroweak symmetry breaking with a light composite scalar resonance.

\section*{Acknowledgments}
This work has been partially supported by a Marie Curie Early Initial Training Network Fellowship of the European Community's Seventh Framework Programme under contract number (PITN-GA-2008-237920-UNILHC) and the MNiSzW scientific research grant N202 103838 (2010 - 2012).

\end{document}